\documentclass[10pt]{article}
\usepackage{a4wide}                      
\usepackage{epsf}                        
\usepackage{latexsym}                    
\usepackage{amsfonts}                    
\usepackage{amssymb}                     
\usepackage{slashed}                     
\usepackage{w-greek}                     
\usepackage{mathtools}
\usepackage{cite}

\usepackage[unicode=true,colorlinks=true, urlcolor=blue, linkcolor=blue, citecolor=blue]{hyperref}

\def\be{\begin{equation}}
\def\ee{\end{equation}}
\def\bea{\begin{eqnarray}}
\def\eea{\end{eqnarray}}
\def\nn{\nonumber \\}
\def\nnw{\nonumber \\ [.2cm]}
\def\vsp#1{\vspace*{#1}}
\def\hsp#1{\hspace*{#1}}

\def\part{\partial}
\def\tfrac#1#2{{\textstyle{\frac{#1}{#2}}}}

\def\cR{{\cal R}}

\def\oGamma{{\mathring \Gamma}}
\def\bGamma{{\bar \Gamma}}
\def\Upgamma{\upgamma\, }

\def\sqrtg{\sqrt{|g|}}

\def\bGamma{\bar{\Gamma}}

\def\mn{{\mu\nu}}
\def\mnr{{\mu\nu\rho}}

\def\inum{\mathrm{i}}
\def\enum{\mathrm{e}}
\def\dder{\mathrm{d}}

\def\makeatletter{\catcode`\@=11}
\makeatletter
\def\mathbox#1{\hbox{$\m@th#1$}}%
\def\math@ccstyles#1#2#3#4#5#6#7{{\leavevmode
      \setbox0\mathbox{#6#7}%
      \setbox2\mathbox{#4#5}%
      \dimen@ #3%
      \baselineskip\z@\lineskiplimit#1\lineskip\z@
      \vbox{\ialign{##\crcr
             \hfil \kern #2\box2 \hfil\crcr
             \noalign{\kern\dimen@}%
             \hfil\box0\hfil\crcr}}}}
\def\mathaccstyles{\math@ccstyles\maxdimen}
\def\maththroughstyles{\math@ccstyles{-\maxdimen}}
\def\unity%
 {\maththroughstyles{.45\ht0}\z@\displaystyle {\mathchar"006C}\displaystyle 1}
\begin{document}

\rightline{\today}
{~}
\vspace{.4truecm}

\centerline{\Large \bf Projective symmetries and induced electromagnetism}
\vspace{.3truecm}
\centerline{\Large \bf in metric-affine gravity}
\vspace{.5truecm}

\centerline{
    {\bf Bert Janssen and}
    {\bf Alejandro Jim\'enez-Cano}\footnote{E-mails: {\tt bjanssen@ugr.es},
                                                     {\tt alejandrojc@ugr.es}}}

\vspace{.4cm}
\centerline{{\it Departamento de F\'isica Te\'orica y del Cosmos and}}
\centerline{{\it Centro Andaluz de F\'isica de Part\'iculas Elementales}}
\centerline{{\it Facultad de Ciencias, Avda Fuentenueva s/n,}}
\centerline{{\it Universidad de Granada, 18071 Granada, Spain}}
\vspace{.5cm}

\centerline{\bf ABSTRACT}
\vspace{.2cm}

\begin{center}
\begin{minipage}{13cm}
We present a framework in which the projective symmetry of the Einstein-Hilbert action in
metric-affine gravity is used to induce an effective coupling between the Dirac lagrangian 
and the Maxwell field. The effective $U(1)$ gauge potential arises as the trace of the
non-metricity tensor $Q_{\mu a}{}^a$ and couples in the appropriate way to the Dirac fields
to in order to allow for local phase shifts. On shell, the obtained theory is
equivalent to Einstein-Cartan-Maxwell theory in presence of Dirac spinors. 
\end{minipage}
\end{center}
\vspace{.4cm} 



\noindent
{{\bf 1. Introduction}}
\vsp{.2cm}

Metric-affine gravity is an extension of the usual (metric) theories of gravity, where the affine
connection $\Gamma_\mn{}^\rho$ is considered as an independent (in general, dynamical) degree
of freedom, whose expression is ultimately determined by its own equation of motion.
In particular, it has been shown \cite{Pons, BJJOSS} that for the $N$-dimensional
Einstein-Hilbert action,
$S=\frac{1}{2\kappa}\int \dder^Nx \sqrtg \cR(\Gamma)$ ($N>2$), the most general expression for the
affine connection is of the form
\be
  \bGamma_\mn{}^\rho \ = \ {\mathring\Gamma}_\mn{}^\rho \ + \ V_\mu \,\delta_\nu^\rho,
  \label{PalatiniGamma}
\ee
where $\oGamma_\mn{}^\rho$ is the Levi-Civita connection and $V_\mu$ an arbitrary vector field.
It turns out, however, that this vector field  $V_\mu$ does not have any physically measurable
influence: both the Einstein equation
and the geodesic equation turn out to be identical to the equations in the metric formalism.
In this sense, the metric and the Palatini formalism are equivalent for the Einstein-Hilbert
action.

In \cite{BJJOSS} it was argued
that the vector field $V_\mu$ is related to the reparametrisation freedom of geodesics:
affine geodesics of the connection $\bGamma_\mn{}^\rho$ turn out to be pre-geodesics of the
Levi-Civita connection $\oGamma_\mn{}^\rho$, through the reparametrisation
\be
\frac{\dder\tau}{\dder\lambda}(\lambda) \ =  \
   \exp  \Bigl[\int_0^\lambda \frac{\dder x^\rho}{\dder\lambda'} V_\rho \, \dder\lambda'\Bigr],
\ee
where $\lambda$ is the affine parameter for the $\bGamma_\mn{}^\rho$ geodesics
and $\tau$ the proper time along the Levi-Civita ones.

In a certain way, the solution (\ref{PalatiniGamma}) reflects the projective symmetry
$\Gamma_\mn{}^\rho \rightarrow \Gamma_\mn{}^\rho + V_\mu \delta_\nu^\rho$, under which the
Einstein-Hilbert is known to be invariant\cite{Eisenh, JS}. Indeed, the Riemann tensor transforms
under the projective transformation as
$\cR_ \mnr{}^\lambda  \rightarrow \cR_ \mnr{}^\lambda \, +  \, F_\mn(V)  \, \delta_\rho^\lambda$
and the Ricci scalar $\cR = g^{\mu\rho}\,\delta_\lambda^\nu\,\cR_\mnr{}^\lambda$ is
easily seen to be invariant.

In more general theories, which include matter terms that couple to the connection, the
Palatini connection (\ref{PalatiniGamma}) might not be a solution and the metric and the
Palatini formalism will in general no longer
be equivalent. However, as we will show, there is a way to restore the symmetry and at
the same time give a physical meaning to the Palatini vector $V_\mu$. The previously
arbitrary vector field will start to play the role of the electromagnetic potential of
Maxwell theory and the projective symmetry of the entire action will be related to
local $U(1)$ transformations. The resulting theory will turn out to be on-shell equivalent
to Einstein-Cartan-Maxwell theory, which describes General Relativity with a torsionful
connection, coupled to the electromagnetic field. Einstein-Cartan theory is considered
\cite{Hehl-book} to be the most likely classical gauge theory of gravity. See for example
\cite{Poplawski} for interesting phenomenology in Einstein-Cartan theory.

We will present our model in the context of Einstein-Dirac theory in the metric-affine
set-up, but it can equally well be done in presence of complex scalar fields. Since
we will be working with Dirac spinors, we will switch to the tangent space description, 
where the metric degrees of freedom are represented by the \emph{Vielbeins} $e^a{}_\mu$,
which are the components of a local orthonormal coframe. The metric in this basis is
then Minkowski, $\eta_{ab} = e^\mu{}_a e^\nu{}_b g_\mn$, in any point of the manifold.
Additionally, the affine connection is substituted by the components of the connection
one-form, $\omega_{\mu a}{}^b$, through the appropriate basis transformation (sometimes
called Vielbein Postulate). We will refer to  $\omega_{\mu a}{}^b$ simply as \emph{connection}
from now on. Note that no antisymmetry in the last two indices is assumed, as the affine
connection is not necessarily metric-compatible.

\vsp{.4cm}
\noindent
{{\bf 2. Spinor covariant derivatives}}
\vsp{.2cm}  

The first issue to address is the construction of the covariant derivative for the
Dirac spinor in the metric-affine context. A popular choice is the natural extension of
the Lorentz covariant derivative to arbitrary (non-symmetric)
connections:\footnote{We use the notation
  $\Upgamma^{ab...c}\equiv \Upgamma^{[a}\Upgamma^{b}...\Upgamma^{c]}$.}
\be
\nabla_\mu \psi \, = \, \partial_\mu \psi
      \, - \, \tfrac{1}{4} \omega_{\mu ab} \Upgamma^{ab}\psi.
\ee
However, as the Lorentz generator $\tfrac{1}{2}\Upgamma^{ab}$ is antisymmetric, only the
antisymmetric part of the connection, $\omega_{\mu [ab]}$, will couple to the spinors. 
Another possibility is to enhance the Lorentz generators to the general product of two
gamma matrices and write a covariant derivative of the form 
\be
\nabla_\mu \psi \, = \, \partial_\mu \psi
      \, - \, \tfrac{1}{4} \omega_{\mu ab} \Upgamma^a \Upgamma^b \psi.
\label{Dpsi1}
\ee 
Notice however that, even though $\omega_{\mu ab} \Upgamma^a \Upgamma^b$  contains more
terms than $\omega_{\mu ab} \Upgamma^{ab}$, most of the symmetric terms are projected out,
due to the anti-commutation relations of the gamma matrices,
$\{\Upgamma^a, \Upgamma^b\} = 2\eta^{ab}$. The difference between the two covariant
derivatives therefore reduces to the trace $2\omega_{\mu ab} \eta^{ab}\psi$, and the expression
(\ref{Dpsi1}) can be written without loss of generality as
\be
\nabla_\mu \psi \, = \, \partial_\mu \psi \, - \, \tfrac{1}{4} \omega_{\mu ab} \Upgamma^{ab} \psi
\, - \, \tfrac{1}{8} Q_{\mu a}{}^a \psi,
\ee
where we have used the definition of the non-metricity tensor
$Q_{\mu ab}\equiv-\nabla_\mu \eta_{ab}$ to relate its trace to the trace of the connection,
$\omega_{\mu a}{}^{a}=\tfrac{1}{2} Q_{\mu a}{}^a$. There is actually no way the spinor can couple
to the traceless symmetric part of the connection. Inspired by \cite{Adak, Hurtley}
(see also \cite{Koivisto, Berredo}), we will
write down a generalised expression for the spinor covariant derivative,
\be
\nabla_\mu \psi \, = \, \partial_\mu \psi \, - \, \tfrac{1}{4} \omega_{\mu ab} \Upgamma^{ab} \psi
    \, - \, k Q_{\mu a}{}^a \psi,
\label{spinorderiv}
\ee
(with $k$ in principle an arbitrary complex parameter), try to identify the gauge group 
this connection corresponds to and determine its dynamics through the first-order
formalism. 

It is not difficult to see that if $k$ is real-valued, the extra term in (\ref{spinorderiv})
is the gauge term for local rescalings $\psi \rightarrow \enum^{k\Omega} \psi$, provided that
the trace of the non-metricity transforms
as $Q_{\mu a}{}^a \rightarrow Q_{\mu a}{}^a + \partial_\mu \Omega$ \cite{TW}.
On the other hand, for
purely imaginary values of $k = \inum e$, (\ref{spinorderiv}) is a covariant derivative
for local $U(1)$ transformations, $\psi \rightarrow \enum^{\inum e\Lambda} \psi$, provided
that $Q$ transforms as $Q_{\mu a}{}^a \rightarrow Q_{\mu a}{}^a + \partial_\mu \Lambda$. 

Of course the trace of the non-metricity is not an independent field, but a part of the
full connection $\omega_{\mu a}{}^b$. We should therefore embed the transformation of
$Q_{\mu a}{}^a$ in a general transformation rule for $\omega_{\mu a}{}^b$. The simplest way
of doing this are the transformation rules, respectively for a dilatation and a $U(1)$
transformation,
\bea
 \omega_{\mu a}{}^b \,  \rightarrow  \, \omega_{\mu a}{}^b \, + \,
                            \frac{1}{2N}\, \partial_\mu \Omega \, \delta_a^b, \hsp{2cm}
 \omega_{\mu a}{}^b \, \rightarrow  \, \omega_{\mu a}{}^b \, + \,
 \frac{1}{2N}\, \partial_\mu \Lambda \, \delta_a^b,
 \label{projsymm}
\eea
which turn out to be projective transformations under which the Einstein-Hilbert action is
invariant \cite{JS, TW, Pons}. For the rest of this letter, we will focus on the $U(1)$ case
and consider only the covariant derivative
\be
\nabla_\mu \psi \, = \, \partial_\mu \psi \,- \, \tfrac{1}{4} \omega_{\mu ab} \Upgamma^{ab} \psi
\, - \, \inum e Q_{\mu a}{}^a \psi.
\label{spinorderiv2}
\ee
together with the transformation rule (\ref{projsymm}b).

\vsp{.4cm}
\noindent
{{\bf 3. Induced electromagnetism from non-metricity}}
\vsp{.2cm}

We are now in the position to write down an action for Einstein-Dirac theory that is
invariant under the combined local $U(1)$ phase shifts and  projective transformation
(\ref{projsymm}b):
\be
S  =\int \dder^Nx \, |e| \, \Bigl[\tfrac{1}{2\kappa} \cR(\omega)
                  \ - \ \tfrac{1}{4} F_\mn(Q) F^\mn(Q) 
                  \ + \ \tfrac{\inum\hbar}{2}(\bar \psi \Upgamma^\mu \nabla_\mu \psi
\ + \ \nabla_\mu \bar\psi  \Upgamma^\mu \psi )
\ - \ m\bar\psi\psi\Bigr].
\label{EDP}
\ee
Here $\cR(\omega)$ is the Ricci scalar of the full connection,
$\cR(\omega) = e^\nu{}_b e^\mu{}_c \eta^{ac} \cR_{\mn a}{}^b(\omega)$ with
$\cR_{\mn a}{}^b (\omega)= 2\partial_{[\mu} \omega_{\nu] a}{}^b
- 2\omega_{[\mu |a}{}^c \omega_{|\nu] c}{}^b$ and $F_\mn(Q) = 2\partial_{[\mu} Q_{\nu]a}{}^a$.
Note that we have added explicitly a gauge invariant kinetic term for the trace
of the non-metricity, since these degrees of freedom do not appear in $\cR(\omega)$,
as a consequence of the projective symmetry. This term was also introduced in \cite{TW},
though without explicitly relating it to the $U(1)$ symmetry of the matter
action.\footnote{Including this term is equivalent to adding the quadratic curvature
term $\cR_{\mn a}{}^a\cR^\mn{}_b{}^b$ (see for example \cite{HM}). \label{RR}}

It is important to realise that the only degrees of freedom of the action (\ref{EDP}) are the 
Vielbeins $e^a{}_\mu$, the connection $\omega_{\mu a}{}^b$ and the Dirac spinor $\psi$. 
In particular, at this stage $Q_{\mu a}{}^a$ is not an independent field and there is no a 
priori reason to think of $Q_{\mu a}{}^a$ as the Maxwell potential. 
The dynamics of $Q_{\mu a}{}^a$ should in fact be fully derived from the dynamics of
$\omega_{\mu a}{}^b$.
 
The equation of motion of the connection,
\bea
0 \ = \ \frac{\kappa}{|e|}\,e^b{}_\sigma\, e^\rho{}_a\,\frac{\delta S }{\delta \omega_{\mu a}{}^b}
   &=& \tfrac{1}{2}T_{\lambda\sigma}{}^{\mu}g^{\rho\lambda} 
      \ - \ \delta_{\sigma}^{[\mu}Q_{\lambda}{}^{\lambda]\rho}
      \ + \ \delta_{\sigma}^{[\mu}g^{\lambda]\rho}
               \Bigl(\tfrac{1}{2}Q_{\lambda\tau}{}^{\tau}-T_{\lambda\tau}{}^{\tau}\Bigr) \nn
    && + \, \kappa \left[2\left(\mathring{\nabla}_\nu F^{\nu\mu}(Q)
                  \, - \, e\hbar\,\bar{\psi}\Upgamma^{\mu}\psi\right)\delta_{\sigma}^{\rho}
      \ - \ \tfrac{\inum\hbar}{4}\,\bar{\psi}\Upgamma^{\mu\rho}{}_{\sigma}\psi\right],
    \label{eomw}
\eea
can be solved in full generality. Indeed, taking the $\delta_\rho^\sigma$ trace, the equation
reduces to
\be
\mathring{\nabla}_\nu F^{\nu\mu}(Q) \, = \, e\hbar\,\bar{\psi}\Upgamma^{\mu}\psi,
\label{eqMax}
\ee
with $\mathring\nabla$ the Levi-Civita covariant derivative. This is clearly the Maxwell
equation with the vector Dirac current as source term, where the trace of the non-metricity
is playing the role of the Maxwell potential, $Q_{\mu a}{}^{a}\equiv A_\mu$
(see also \cite{TW}).
Once \eqref{eqMax} is taken into account, the equation of motion (\ref{eomw}) simplifies
considerably: if we define 
$S_{\mu a}{}^{b} \equiv \tfrac{\inum \hbar}{4}\bar{\psi}\Upgamma_{\mu a}{}^{b}\psi$,
then the most general solution of the traceless part of (\ref{eomw}) is given by
\cite{Pons, TFM}
\be
\omega_{\mu a}{}^{b} \, =\, \mathring{\omega}_{\mu a}{}^{b} \, + \, V_{\mu}\,\delta_{a}^{b}
\, + \,\kappa S_{\mu a}{}^{b},
\label{wsol}
\ee
where $\mathring{\omega}_{\mu a}{}^{b}$ is the Levi-Civita connection in the anholonomic
frame and $V_{\mu}$ is a one-form, which is basically the vector field encountered in
(\ref{PalatiniGamma}). While for the Einstein-Hilbert action the vector field $V_\mu$
was completely arbitrary and void of physical content, in this context it takes up
the role of the Maxwell potential $A_\mu$. Indeed, it is straightforward to see that
$V_\mu$ represents the trace $Q_{\mu a}{}^{a}=2NV_{\mu}$ of the non-metricity
$Q_{\mu ab} =  2V_{\mu} \eta_{ab}$ of the solution (\ref{wsol}) and hence not only is its
dynamics dictated by the Maxwell equation (\ref{eqMax}), but also the projective
symmetry (\ref{projsymm}) makes it behave as a $U(1)$ gauge field. For future reference,
we note that the torsion of the solution (\ref{wsol}) is given by
$T_{\mn}{}^\rho = 2 V_{[\mu}\delta_{\nu]}^\rho + 2\kappa S_\mn{}^\rho$.

On the other hand, the equations of motion for Dirac spinor $\psi$ can be easily computed.
Upon substituting (\ref{wsol}), its takes the form of the inhomogeneous Dirac equation coupled
to the electromagnetic field, with $S_{\mu a}{}^{b}$ acting as its source term:
\be
\inum\hbar\, \Upgamma^{\mu}\Bigl[\mathring{\nabla}{}_{\mu} - \inum eA_\mu\Bigr] \psi \ - \ m\psi 
 \ =  \ \tfrac{1}{4}\inum\hbar\kappa \, \Upgamma^{\mu\nu\rho} S_{\mu\nu\rho}\,\psi.
\label{Diraceq}
\ee
Note that the torsion and non-metricity terms $Q_{[\lambda\mu]}{}^\lambda + T_{\mu\lambda}{}^\lambda$
which appear in \cite{FR, DOR} cancel in our case, as they are both proportional to $A_\mu$.

Finally, after taking in account the solution (\ref{wsol}), the Vielbein equation of motion
splits into its symmetric and antisymmetric parts,
\bea
\mathring{\cR}_{\mu\nu}-\tfrac{1}{2}g_{\mu\nu}\mathring{\cR} 
 &=&  \tfrac{1}{2}\kappa^2 g_{\mu\nu}S^{\rho\lambda\sigma}S_{\rho\lambda\sigma} 
\ + \ \kappa\,\Bigl[F_{\mu}{}^{\lambda}(A)F_{\nu\lambda}(A)
                \, -\, \tfrac{1}{4}g_{\mu\nu}F_{\lambda\sigma}(A)F^{\lambda\sigma}(A)\Bigr] \nn
&& \hsp{1cm}
-\tfrac{\inum\hbar \kappa}{2}\left[
  \bar{\psi}\Upgamma_{(\nu}\Bigl(\mathring{\nabla}_{\mu)} -\inum e A_{\mu)}\Bigr)\psi
     \, - \, \Bigl(\mathring{\nabla}_{(\mu} +\inum e A_{(\mu}\Bigr)\bar{\psi}\Upgamma_{\nu)}\psi\right],
                    \nnw
\mathring{\nabla}_{\lambda}S_{\mu\nu}{}^{\lambda} 
&=& \tfrac{\inum\hbar}{2}\left[
  \bar{\psi}\Upgamma_{[\nu}\Bigl(\mathring{\nabla}_{\mu]} -\inum e A_{\mu]}\Bigr)\psi
     \, - \, \Bigl(\mathring{\nabla}_{[\mu} +\inum e A_{[\mu}\Bigr)\bar{\psi}\Upgamma_{\nu]}\psi\right],
\label{Einsteineq}
\eea
which play the role of the Einstein equation and an equation that renders the torsion
dynamical. Note that the total energy-momentum tensor on the right-hand side of the
Einstein equation does not only contain the standard contributions of the electromagnetic
and the Dirac field, but also an additional contribution of $S_{\mu a}{}^{b}$,
quadratic in $\kappa$. The set of equations (\ref{eqMax}), (\ref{Diraceq}) and
(\ref{Einsteineq}) are the usual equations of motion of Einstein-Cartan gravity, coupled
to the electromagnetic field and to Dirac spinors.

\vsp{.4cm}
\noindent
{{\bf 4. Discussion}}
\vsp{.2cm}

We have shown that in the framework of metric-affine gravity, the Maxwell field can be
geometrised and be interpreted as part of a more general connection $\omega_{\mu a}{}^{b}$.
In particular, the Maxwell equation is encoded in (the trace of) the equation of motion
of the connection. In order to obtain this it is necessary to relate the $U(1)$ transformation
of the matter fields with the projective transformation of the connection that leaves the
Einstein-Hilbert term, and hence the entire action,  invariant. The equations of motion,
after substituting the solution for the connection, turn out to be equivalent to the
equations of motion of the Einstein-Cartan theory (i.e. the Einstein theory with a
metric-compatible torsionful connection) in the presence of electromagnetism and a Dirac
field.

Actually, it is not that surprising that the action (\ref{EDP}) is on-shell equivalent
to Einstein-Cartan gravity, coupled to the Maxwell and Dirac fields. Consider the
decomposition of the general connection $\omega_{\mu a}{}^{b}$ into its antisymmetric
(and hence metric-compatible), its traceless symmetric and its trace parts,
\be
\omega_{\mu a}{}^{b} \ =\ \bar{\omega}_{\mu a}{}^{b} \ + \ V_{\mu} \,\delta_{a}^{b}
\ + \ \tilde{Q}_{\mu a}{}^{b},
\ee
where  $\bar{\omega}_{\mu}{}^{(ab)}= \tilde{Q}_\mu{}^{[ab]} = \tilde{Q}_{\mu c}{}^{c}=0$.
Under this decomposition, the Ricci scalar $\cR(\omega)$ takes the form
\be
\cR(\omega) \ = \ \bar{\cR}(\bar{\omega}) \ + \ \tilde{Q}_{\mu\nu\lambda}\tilde{Q}^{\nu\mu\lambda}
         \ - \ \tilde{Q}_{\mu}{}^{\mu \lambda}\tilde{Q}^{\nu}{}_{\nu \lambda},
\ee
where $\bar{\cR}(\bar{\omega})$ is the Ricci scalar of  $\bar{\omega}_{\mu ab}$, which 
is a connection in its own right. Note that the trace $V_\mu$ does not contribute to the
Ricci scalar, due to the projective symmetry.

With this decomposition, the action (\ref{EDP}) can be written as
\be
S  =\int \dder^Nx \, |e| \, \Bigl[\tfrac{1}{2\kappa} \bar{\cR}(\bar\omega)
         \, + \, \tfrac{1}{2\kappa}\Bigl(\tilde{Q}_{\mu\nu\lambda}\tilde{Q}^{\nu\mu\lambda}
                 \, - \, \tilde{Q}_{\mu}{}^{\mu \lambda}\tilde{Q}^{\nu}{}_{\nu \lambda} \Bigr)
         \, - \, \tfrac{1}{4} F_\mn(V) F^\mn(V)
         \, + \,  \mathcal{L}_{\mathrm{Dirac}}
                 \Bigr].
\label{ECMDQ}
\ee
In this set-up it is clear  from its kinetic term and its coupling to the Dirac spinors
that the trace of the connection $V_\mu$ already plays the role of the Maxwell potential.
However the action (\ref{ECMDQ}) is not yet equivalent to the standard
Einstein-Cartan-Maxwell-Dirac theory, due to the presence of the traceless symmetric part
of the connection $\tilde{Q}_\mnr$.

The equation of motion (\ref{eomw}) of the original connection $\omega_{\mu a}{}^{b}$ must be
equivalent to the combined equations of motion of its three parts: on the one hand, the
equation for the metric-compatible connection $\bar{\omega}_{\mu a}{}^{b}$ yields the standard
Einstein-Cartan-Dirac solution $\bar{\omega}_{\mu a}{}^{b} = \mathring{\omega}_{\mu a}{}^{b} 
+ \kappa S_{\mu a}{}^{b}$ and the equation for $V_\mu$ gives rise directly to the Maxwell
equation coupled to the spinors. On the other hand, the traceless symmetric part
$\tilde{Q}_\mnr$ appears quadratically, implying its own triviality, 
\be
\tilde{Q}_{\mu\lambda\nu} \ = \ 0.
\ee
So, with $\tilde{Q}_\mnr$ on-shell, we easily see that the action (\ref{ECMDQ}) becomes
the action of the Einstein-Cartan theory coupled to electromagnetism and a Dirac field.
Consequently, the tensor $S_{\mu a}{}^{b}$ turns out to be the spin-density current
associated to the Lorentz connection $\bar{\omega}_{\mu a}{}^{b}$.

Up to a certain point, the philosophy of this construction is similar to the construction
made in \cite{AOR} and references therein, where standard General Relativity with non-linear
matter terms is reproduced from a class of modified metric-affine theories, called
Ricci-based gravities. However we wish to emphasise that our action (\ref{ECMDQ}) does not
belong to this class of Ricci-based gravities, as it contains terms proportional to
non-Ricci-like quadratic curvature invariant 
$\cR_{\mn a}{}^a\cR^\mn{}_b{}^b$, as pointed out in footnote \ref{RR}. This suggests that the
procedure of \cite{AOR} might apply to a more general class of theories.

\vspace{1cm}
\noindent
{\bf Acknowledgements}\\
The authors would like to thank Jos\'e Ignacio Illana, Jos\'e Alberto Orejuela and Javier
Mart\'inez Lizana for useful discussions. This work was partially supported by the Junta
de Andaluc\'ia (FQM101), the Universidad de Granada (PP2015-03) and the Spanish Ministry of
Economy and Competitiveness (FIS2016-78198-P). AJC is supported by a PhD contract of the
program FPU 2015 with reference FPU15/02864 (Spanish Ministry of Economy and Competitiveness).


\end{document}